# BenchLink: An SoC-Based Benchmark for Resilient Communication Links in GPS-Denied Environments

Sidharth Santhi Nivas[1], Prem Sagar Pattanshetty Vasanth Kumar[2], Zhaoxi Zhang[1], Chenzhi Zhao[1], Maxwell McManus[2], Nicholas Mastronarde[2], Elizabeth Serena Bentley[3], George Sklivanitis[4], Dimitris A. Pados[4], and Zhangyu Guan[1]

[1]Department of Computer Science & Engineering, University of Minnesota - Twin Cities, MN 55455, USA

[2]Department of Electrical Engineering, University at Buffalo, NY 14260, USA

[3]U.S. Air Force Research Laboratory (AFRL), NY 13441, USA

[4]Department of Electrical Engineering and Computer Science, Florida Atlantic University, FL 33431, USA

Email: {santh055, zha00410, zhao2305, zguan}@umn.edu, {premsaga, memcmanu, nmastron}@buffalo.edu, elizabeth.bentley.3@us.af.mil, {gsklivanitis, dpados}@fau.edu

***Abstract*—Accurate timing and synchronization, typically enabled by GPS, are essential for modern wireless communication systems. However, many emerging applications must operate in GPS-denied environments where signals are unreliable or disrupted, resulting in oscillator drift and carrier frequency impairments. To address these challenges, we present BenchLink, a System-on-Chip (SoC)-based benchmark for resilient communication links that functions without GPS and supports adaptive pilot density and modulation. Unlike traditional General Purpose Processor (GPP)-based software-defined radios (e.g. USRPs), the SoC-based design allows for more precise latency control. We implement and evaluate BenchLink on Zynq UltraScale+ MPSoCs, and demonstrate its effectiveness in both ground and aerial environments. A comprehensive dataset has also been collected under various conditions. We have made both the SoC-based link design and dataset available to the wireless community. BenchLink is expected to facilitate future research on data-driven link adaptation, resilient synchronization in GPS-denied scenarios, and emerging applications that require precise latency control, such as integrated radar sensing and communication.**

***Index Terms*—GPS-Denied Communication, System-on-Chip (SoC), Field-Programmable Gate Array (FPGA), Software-Defined Radio (SDR).**

## I. Introduction

Global Positioning System (GPS) technology has long been a key component of modern wireless communication systems, providing accurate timing and positioning information [1], [2]. For example, cellular networks rely heavily on GPS to maintain synchronization among base stations, allowing seamless handovers, supporting time division duplexing (TDD), and facilitating advanced techniques such as Coordinated Multi-Point (CoMP) [3], heterogeneous edge networks [4], and hybrid aerial-ground networking [5]. However, GPS signals can be unreliable, unavailable, or intentionally disrupted in many emerging applications, such as vehicle networks in the field and unmanned aerial vehicles (UAVs), where they are particularly vulnerable to interference and spoofing [6], [7]. Similarly, in civilian advanced air mobility (AAM) and urban vehicular networks, GPS coverage may be obstructed by urban canyons, tunnels, or indoor environments [8], [9]. In such scenarios, the sole reliance on GPS for synchronization becomes impractical and jeopardizes the stability and performance of the communication system.

A fundamental challenge in these communication scenarios is maintaining synchronization between the local oscillators of the transmitter and receiver without GPS, as environmental and operational factors cause oscillator drift, leading to carrier frequency offset (CFO) and rotation of the constellation points over time [10]. This challenge is further exacerbated by mobility, which introduces rapid fluctuations in signal strength and phase due to sudden changes in position, orientation, vibrations, and dynamic multipath propagation [11], [12].

To mitigate these challenges, our objective in this work is to provide a programmable benchmark design for wireless communication links based on System-on-Chip (SoC) technologies that can operate effectively in *real-world environments* without relying on GPS-based clock synchronization. Compared to GPP-based software-defined radios (SDRs), such as Universal Software Radio Peripheral (USRP), SoC solutions, which integrate Field-Programmable Gate Arrays (FPGAs), can offer significantly lower and more predictable latency. This characteristic is particularly advantageous in scenarios where precise latency control and low-latency signal processing are critical, such as integrated sensing and communications (ISAC) [13]. To the best of our knowledge, there is currently no widely adopted open-source SoC-based benchmark for programmable wireless communications in GPS-denied environments that is both practical and readily deployable in real-world scenarios.

Specifically, we focus our design on enabling the programmability of the link in adapting to over-the-air (OTA) radio frequency (RF) impairments, particularly phase and frequency offsets caused by clock drift in GPS-denied environments. In current practice, such impairments are typically

This work was supported in part by the National Science Foundation (NSF) under Grant SWIFT-2229563 and CNS-2450418, and the U.S. Air Force Research Laboratory under Contracts FA8750-21-F-1012, FA8750-20-C1021 and FA8750-25-1-1000.

Distribution A. Approved for public release: Distribution Unlimited: AFRL-2025-4300 on 27 Aug 2025.

The BenchLink source code and dataset are available at: https://wingslabwireless.github.io/opensource.html#benchlink

mitigated in two stages [14]. First, a coarse frequency correction is performed using a pilot training sequence at the start of each frame, exploiting its strong autocorrelation properties to estimate and compensate for the initial offset. The residual frequency errors and constant phase offsets are then corrected at the packet level using pilot sequences known *a priori* to both the transmitter and receiver. However, embedding pilot sequences within data packets introduces a fundamental trade-off: increasing the number of pilots improves channel estimation and tracking in dynamic environments but consumes resources that could otherwise carry user data. Conversely, reducing pilot density improves throughput, but risks reliability degradation due to uncorrected carrier-frequency offsets and phase noise, especially in high-order modulations where constellation points are more tightly packed.

**Contributions.** The main contributions of this paper are as follows:

- *SoC-based design and implementation of BenchLink:* We present a reconfigurable communication link on the Zynq UltraScale+ MPSoC that supports adaptive pilot density and modulation for GPS-denied environments. The design follows a hardware/software co-design approach, with the ARM processing system managing RF control and higher-layer protocols, while the FPGA logic accelerates PHY processing to deliver low-latency, deterministic performance. This architecture enables real-time link adaptation and systematically explores the trade-off between pilot overhead and modulation order to ensure robust communication in dynamic scenarios.
- *Experimental evaluation in diverse environments:* We conduct a comprehensive experimental study of the designed link, focusing on the impact of residual frequency offset caused by channel dynamics and clock drift between SoC devices in GPS-denied settings. The evaluation uses controlled burst transmissions with varied packet configurations in ground-to-ground (G2G) and air-to-air (A2A) UAV testbeds under real-world conditions.
- *Dataset for wireless research:* We collect a comprehensive dataset using BenchLink in indoor, outdoor G2G, and A2A scenarios. The dataset spans various physical layer configurations to capture residual frequency offset, goodput, and link quality metrics under a range of channel conditions. All recordings are paired with JSON files that conform to the Signal Metadata Format (SigMF).

The developed SoC-based link can enable a wide range of research topics, including testing AI and machine learning algorithms for real-world link adaptation and dynamic resource allocation, exploring integration with low-latency sensing and control functions, and prototyping ISAC systems under mobility constraints, among others.

The remainder of this paper is organized as follows. Section II reviews related work. In Section III we describe the BenchLink frame structure, and in Section IV we present the SoC-based link implementation. Section V describes the development of the testbed and the experimental evaluation, and Section VI describes the collected dataset. Finally, we discuss limitations and future work in Section VII and draw the main conclusions in Section VIII.

## II. Related Work

Recent studies have investigated synchronization techniques for wireless systems operating in GPS-denied environments, exploring both physical-layer signal processing and network-level coordination. For example, Brunner *et al.* [14] introduce a bistatic OFDM ISAC system with an OTA synchronization process that corrects timing, carrier frequency, and sampling frequency offsets using preamble and pilot symbols. The experimental setup generates signals on an onboard computer, with a Zynq ZCU111 SoC handling digital up-conversion and pass-band processing. In [15], Savaux *et al.* present two Chirp Spread Spectrum (CSS) receiver designs: a non-coherent receiver that estimates timing and frequency using a mixed up/down chirp preamble with energy-based chirp decoding, and a coherent receiver that tracks carrier phase via adaptive loops to enable PSK-CSS demodulation at higher data rates. Jin *et al.* [11] propose a self-time synchronization (STS) scheme for dynamic multi-UAV networks that takes advantage of Doppler-induced frequency offsets to improve clock skew and offset estimation without external references. A fully distributed time synchronization method is developed by Liu *et al.* [16] for UAV formation networks, where the signal phases are aligned via local broadcast and neighbor clock averaging.

Although these studies provide valuable insights into synchronization in GPS-denied environments, they do not investigate the impact of reconfigurable frame structures for mitigating clock drift. Moreover, most of the above studies, along with similar works reported in the literature, are limited to simulations [17], [18] or controlled indoor experiments [19] and their reported performance does not always translate directly to practical deployments [20]–[22]. *The main novelty of this work is the development of an open-source programmable SoC-based wireless link benchmark for real-world GPS-denied environments, leveraging MPSoC hardware/software co-design with FPGA-accelerated PHY processing and adaptive frame structures to achieve low latency and robust operation under clock drift-induced impairments.*

## III. Frame Design for SoC CFO Mitigation

We develop the configurable link on a Multiprocessor System-on-Chip (MPSoC) platform. As mentioned in Section I, we choose MPSoC because it provides more explicit and predictable latency control, enabling the designed link to support low-latency applications such as spectrum sensing [23], ISAC [13], and scenarios requiring precise latency control, such as Positioning, Navigation, and Timing (PNT) [24] in GPS-denied environments. Next, before describing the programmability design of the link, we briefly introduce the hardware and software architecture of the MPSoC platform.

### A. Software–Hardware Partitioning and Co-Design

In this work, we use the Zynq UltraScale+ ZCU102 board as the target MPSoC platform. Figure 1 shows the main components of the MPSoC platform and its interfaces with the northbridge Onboard Computer and the southbridge RF Front-End. The Onboard Computer (an Intel NUC in our case) runs the upper layers of the protocol stack, such as TCP/IP or a custom-designed protocol stack, along with cross-layer optimization functionalities. Data traffic, generated using a live video stream from a webcam connected to the NUC, is fed to the PHY-layer interface. The data and control parameters are then sent to the MPSoC platform via the northbridge interface, which uses Ethernet and is powered by the *libiio* library [25].

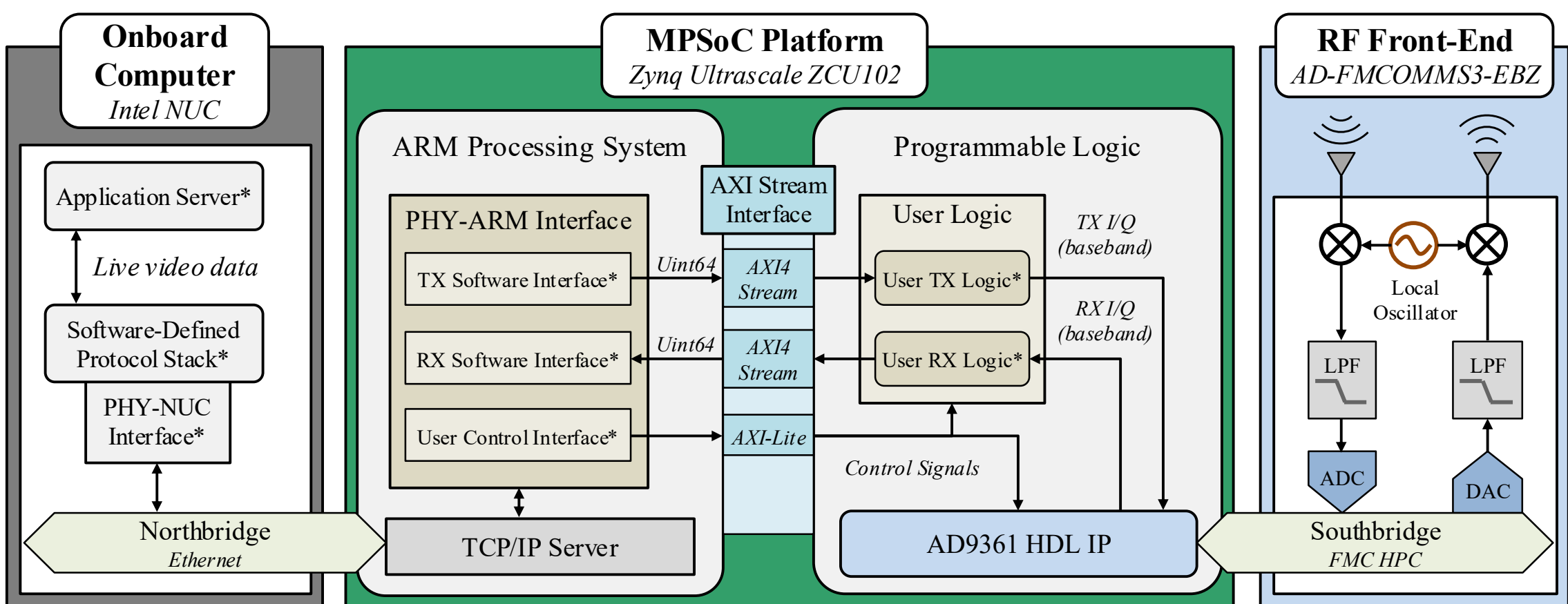

Fig. 1: Software–hardware partitioning and co-design architecture of BenchLink. Components marked with * indicate modules custom designed and implemented as part of this work.

Unlike SDRs that rely on GPPs for baseband signal processing, such as USRPs [26], MPSoCs integrate heterogeneous computing platforms (HCPs), combining processor cores, FPGA fabric, and high-speed interconnects to merge software-programmable processing systems with FPGA-based programmable logic. This architecture enables the FPGA fabric to handle computationally intensive physical layer tasks, such as high-speed baseband signal processing, while the processor manages coordination and higher-level communication protocols.

Specifically, as shown in Fig. 1, the Zynq UltraScale+ ZCU102 MPSoC platform consists of two subsystems: the ARM Processing System (PS) and the Programmable Logic (PL). The PS configures the AD-FMCOMMS3-EBZ RF front-end board by setting key parameters such as center frequency and gain, while the PL handles baseband signal processing. For transmission, the processed baseband I/Q samples are sent to the RF front-end board, which up-converts them to RF signals. On the receiving side, the RF front-end board down-converts the RF signal and sends the baseband samples back to the FPGA for processing. The data then flow through the MPSoC PHY-ARM interface, into the PHY-NUC interface of the Onboard Computer through the Ethernet link, and ultimately up to the upper layers of the protocol stack.

### *B. Frame Design With Configurable CFO Mitigation*

As discussed in Section I, our goal is to design an SoC-based benchmark link whose parameters can be dynamically reconfigured to adapt to random CFO between communicating nodes and time-varying channel conditions. As illustrated in Fig. 1, the carrier frequency is generated by the local oscillator (LO) of the RF front-end interfaced with the MPSoC Platform. Any mismatch or instability between the LOs of the communicating nodes manifests as CFO, while temperature variations and independent reference clocks contribute to gradual drift over time. Although the MPSoC's digital baseband processing is synchronized to its internal clock, this clock is not phase-locked to the RF front-end LO, making the LO the dominant source of CFO and the resulting phase rotation.

Such distortions are typically mitigated at the frame level using autocorrelation-based estimation with fixed preamble sequences. However, this approach can prove insufficient for non-coherent reception, where the transmitter and receiver do not have access to a shared reference clock, such as GPS-provided pulse-per-second (PPS) signals. In BenchLink, in addition to the frame-level coarse CFO correction, we introduce an adaptive fine-tuning mechanism that embeds pilot symbols within configurable subframes to refine channel estimation and compensate for magnitude and phase distortions, as well as the residual frequency offsets remaining after coarse correction.

*1) Coarse Frequency Offset Correction:* Let $\delta_n$ denote the resulting frequency distortion at the $n$-th sample, which can be expressed as $\delta_n = e^{-j(2\pi \Delta f n T_{\texttt{sp}} + \theta_{\texttt{in}})}$, where $T_{\texttt{sp}}$ is the sampling period and $\theta_{\texttt{in}}$ and $\Delta f$ are the initial phase mismatch and the CFO estimated at run time. This distortion appears as a gradual phase rotation of the received baseband signal over time.

To estimate $\Delta f$, the receiver first detects the start of the fixed preamble sequence in the received complex baseband symbols, denoted $x[n]$, by calculating the autocorrelation $C[n]$ as

$$C[n] = \sum_{k=n-M+1}^{n} x[k]\, x^*[k-M], \tag{1}$$

where $M$ is the length of the preamble sequence. To detect the start of a preamble sequence, the receiver calculates a decision metric $\rho[n]$ for each window of $M$ symbols, defined as $\rho[n] = \frac{|C[n]|}{P[n]}$, where $P[n] = \sum_{k=n-M+1}^{n} |x[k]|^2$ is the average energy over the window of the observed symbol. The start of a preamble sequence is then detected at the symbol index for which $\rho[n]$ exceeds a programmable decision threshold.

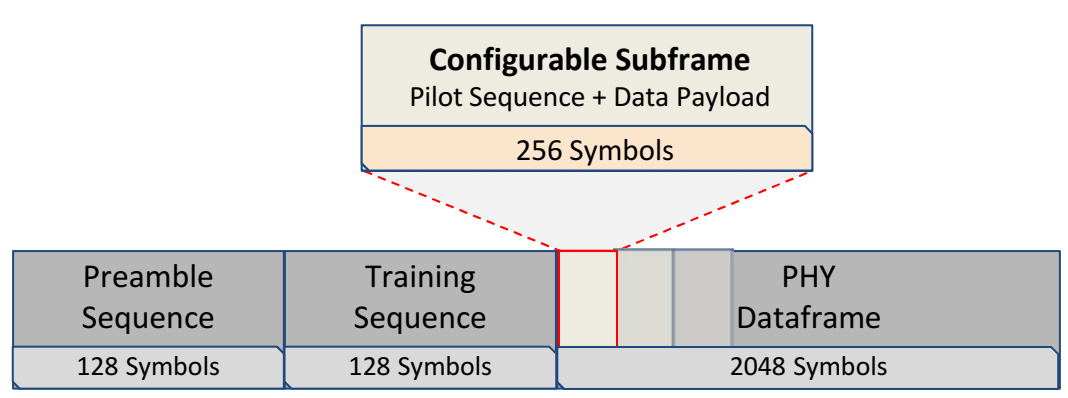

Fig. 2: Configurable Frame Structure of BenchLink.

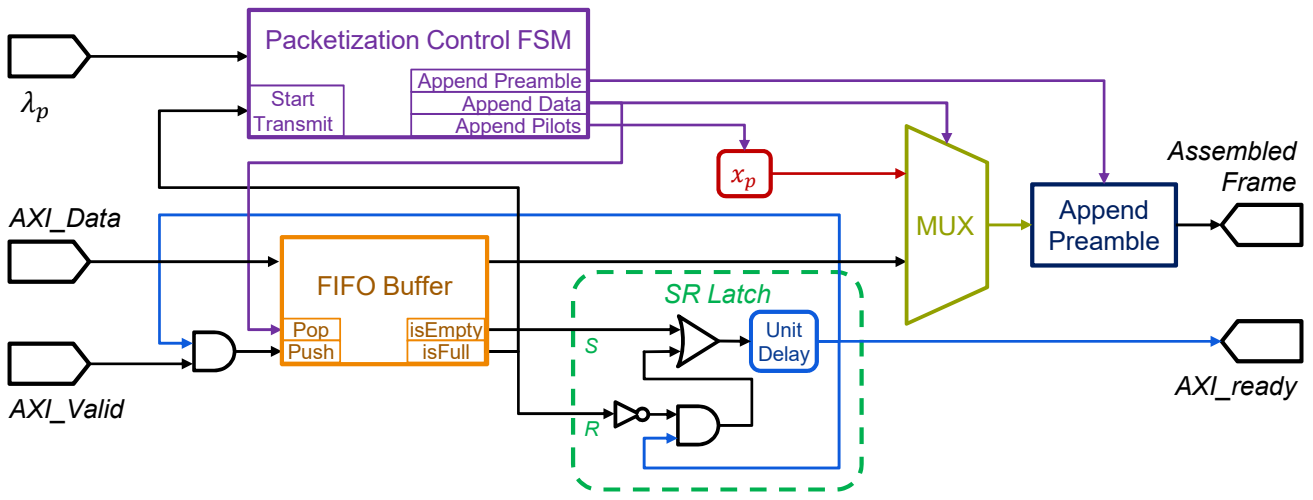

Fig. 3: Dynamic frame assembly for data transmission.

The phase of the associated peak in $C[n]$ at this index, denoted $\angle C_{peak}$, describes the phase deviation experienced by the preamble symbols over a known sample delay, denoted $\varDelta t$. Since this phase deviation accumulates linearly with frequency offset, the normalized CFO can be estimated as $\varDelta f_{\text{est}} = \frac{\angle C_{\text{peak}}}{2\pi \varDelta t}$.

The value of $\varDelta f_{\text{est}}$ is then used to drive a Numerically Controlled Oscillator (NCO), which generates a correction waveform at frequency $-\varDelta f_{\text{est}}$. The correction waveform is finally used to compensate for the received signal, which can be expressed as $y[n] = x[n]\, e^{-j2\pi \varDelta f_{\text{est}} n T_{\text{sp}}}$, effectively removing the bulk phase rotation prior to subsequent fine-grained channel tracking.

*2) Configurable Refined CFO Mitigation:* After coarse frequency correction is applied, the residual frequency offset is typically compensated once per subframe before the next frame undergoes coarse correction. The residual frequency after coarse CFO correction is expressed as $\varDelta\phi = \varDelta f - \varDelta f_{\text{est}}$, where $\varDelta\phi$ captures the residual frequency offset caused by local oscillator drift and channel-induced impairments. This offset introduces phase rotation across the samples, which is typically corrected using pilot sequences known to both the transmitter and receiver and embedded within each subframe. We consider the delay spread of the channel $T_D$ to be much shorter than the sampling interval $T_{\text{sp}}$. As a result, inter-symbol interference due to multipath fading is negligible, and the channel can be modeled as a block fading, single-tap channel [27]. In this case, a single complex coefficient $H$ is sufficient to capture the combined effects of channel fading and $\varDelta\phi$. This coefficient is estimated by correlating the coarse-frequency-corrected received signal $y[n]$ with the conjugate of the time-reversed pilot sequence $x_p[n]$, as

$$H = \frac{1}{N_p} \sum_{n=0}^{N_p - 1} y[n] \cdot x_p^*[N_p - n - 1], \tag{2}$$

where $N_p$ represents the length of the pilot sequence.

In a block fading scenario, the subframe or payload length is typically chosen based on the coherence time $T_c$, under the assumption that the channel remains relatively stable within this interval. However, wireless channels can exhibit widely varying fading characteristics depending on the environment, carrier frequency, hardware nonlinearities, and the relative orientation and position of the transmitter and receiver nodes. Consequently, the channel coherence time $T_c$ can differ significantly across different scenarios. For example, ground-to-ground (G2G) links in static environments often exhibit slow fading and large $T_c$, allowing less frequent channel corrections. In contrast, UAV-involved communication links are more susceptible to Doppler effects or platform jitter, resulting in shorter coherence intervals that may require more frequent channel tracking and correction [28]–[30]. A rigid frame structure optimized for one deployment configuration could be insufficient for another. To address this variability, we introduce a configurable frame structure as shown in Fig. 2 that allows the number of pilot repetitions per packet, denoted $\lambda_p$ to be reconfigured based on the estimated coherence properties of the link.

By adjusting $\lambda_p$, the system adapts the resolution at which temporal channel dynamics are estimated, using more frequent pilot insertion for short $T_c$ and less frequent pilot insertion for long $T_c$. For a payload length of $L$ symbols, pilots are inserted every $\frac{L}{\lambda_p}$ symbols, enabling continuous tracking and correction of the combined RF impairments caused by the channel and the residual frequency offset due to local oscillator drift.

## IV. SOC-BASED TX/RX CHAIN IMPLEMENTATION

As discussed in Sec. I, SoC-based designs with integrated FPGAs can enable deterministic, low-latency processing by implementing time-critical PHY functions directly in hardware. A main challenge with SoC systems is that they impose stricter timing constraints and reduced flexibility, often requiring time-consuming bitstream resynthesis even for minor design changes. To support dynamic frame structure reconfiguration, we implement an asynchronous FIFO buffer and a programmable frame-building state machine that serves to assemble packets based on the value of $\lambda_p$, as shown in Fig. 3. Once the buffer size reaches a configured threshold, the frame-building state machine is activated, which coordinates the timing between different modules responsible for the appending of preamble sequences and interleaving pilot sequences within data symbols according to the programmed $\lambda_p$ for controlled burst transmission. The value of $\lambda_p$ is kept programmable via an AXI-Lite interface, enabling system reconfiguration for diverse channel conditions. On the receiver side, a corresponding frame detection subsystem, shown in Fig. 4, uses a state machine to deconstruct incoming packets and identify pilot sequences based on the configuration.

The transmitter-receiver chain design is shown in Fig. 5. On the transmitter side, data bytes from the AXI4-Stream interface are buffered and dynamically packetized according to the configured pilot sequences. The payload is further appended with necessary preambles and mapped to QAM symbols. To mitigate Inter Symbol Interference (ISI), the samples are then pulse-shaped using a Root Raised Cosine (RRC) filter. Subsequently, the pulse-shaped signal is up-converted and transmitted via the RF front-end. At the receiver, the RF front-end performs down-conversion, after which the signal undergoes automatic gain control (AGC) and matched filtering using a corresponding RRC filter. Frequency and phase offsets are subsequently estimated and corrected on the basis of configured parameters. Finally, the signal is demodulated into data symbols and transferred to the ARM processing system through an AXI4-Stream interface.

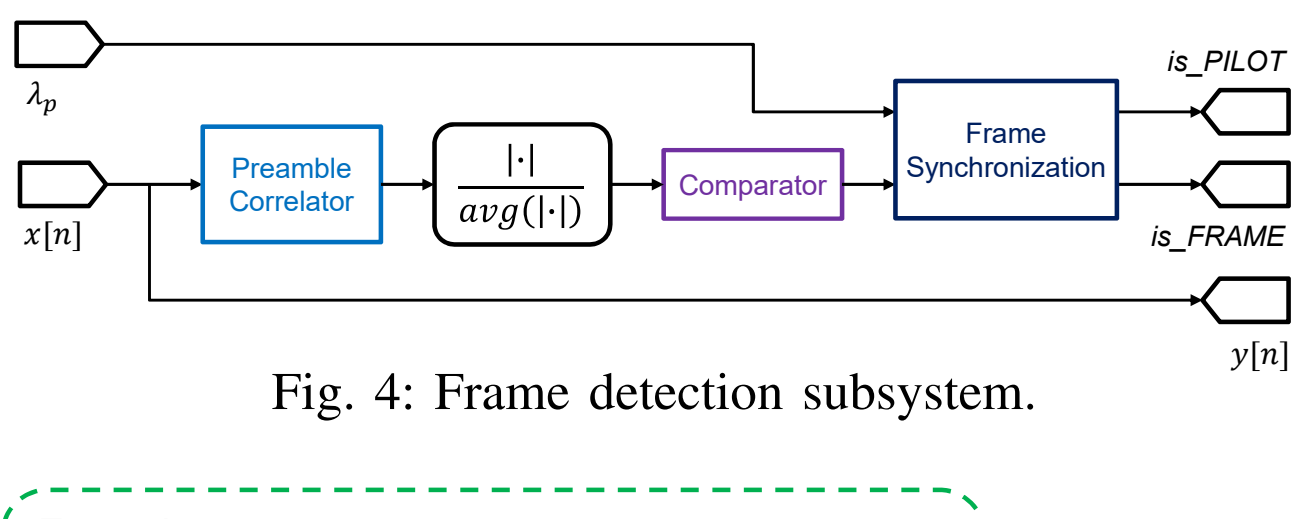

Fig. 4: Frame detection subsystem.

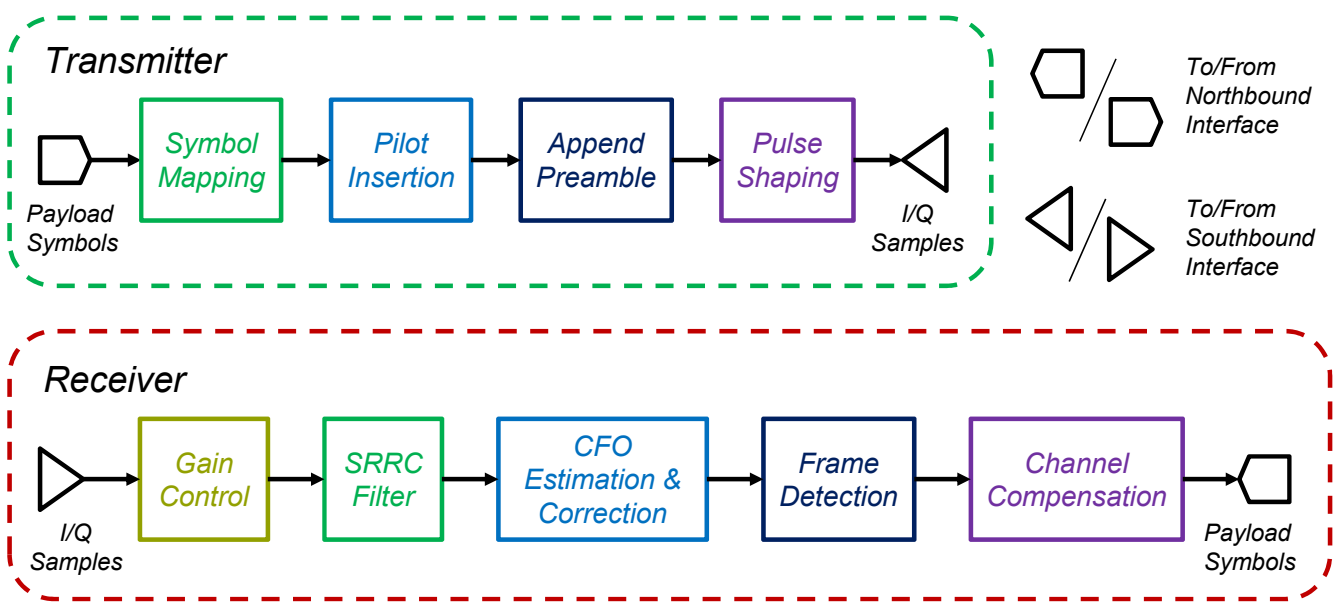

Fig. 5: Transmitter and receiver chains.

**Data Generation**: Traffic data, e.g., live video from a webcam connected to the onboard Intel NUC, is captured, encoded and transmitted over a UDP socket to the physical layer interface running on the ARM processor. This stream is then forwarded to a FIFO buffer implemented in the programmable logic using the AXI4-Stream protocol. The AXI4-Stream interface consists of three primary signals: `TDATA` for data, `TVALID` to indicate when data is valid, and `TREADY` asserted by the programmable logic to apply backpressure and manage flow control. Data transfer occurs only when both `TVALID` and `TREADY` are high, ensuring reliable streaming. The received data are stored in the FIFO, where it is later used for packet assembly and transmission during the next scheduled burst.

**Packet Assembly & Symbol Mapping**: The HDL-based packet synchronization module, shown in Fig. 3, is governed by a Moore state machine that ensures the correct alignment and sequencing of the packetization process. Based on the configured number of pilot sequences per packet, $\lambda_p$, the module interleaves the data payload with pilot symbols retrieved from a lookup table. The assembled packet is then prepended with a preamble for frame synchronization and a training sequence for coarse CFO correction. Following this, the byte

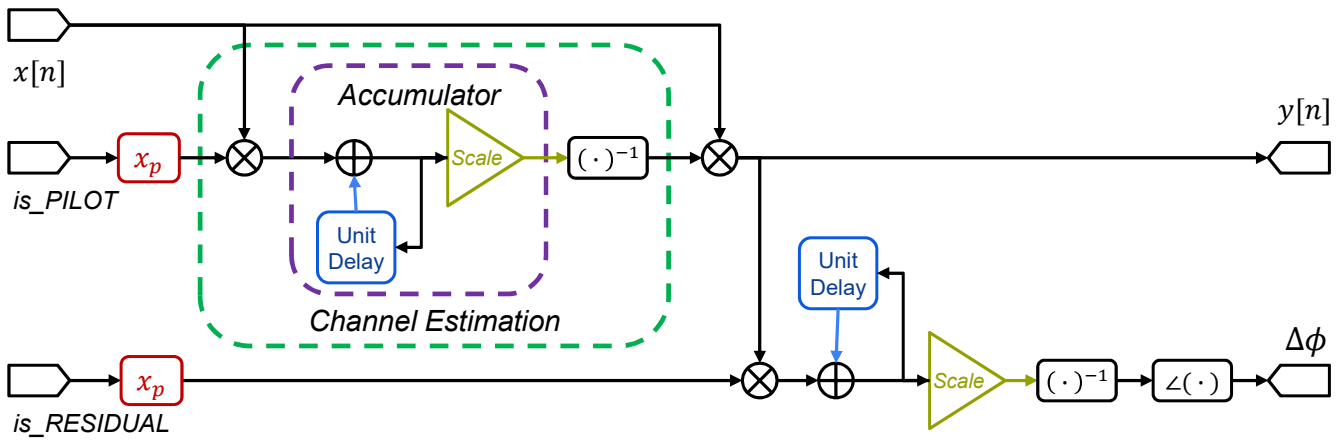

Fig. 6: Channel equalization and estimation of residual CFO.

stream is mapped to complex I/Q symbols using configurable modulation schemes.

**Pulse Shaping & RF Transmission**: To mitigate ISI, pulse shaping is applied to the sample sequence. The signal is first upsampled by an integer interpolation factor $l$ and then convolved with a Square-Root Raised-Cosine (SRRC) pulse-shaping filter, realized as a Finite Impulse Response (FIR) filter. Finally, data are transmitted through the RF front-end, which performs digital-to-analog conversion, upconversion to the desired carrier frequency, and OTA transmission as shown in Fig. 1.

**Signal Reception, Conditioning & CFO Correction**: The received RF signal is downconverted to baseband and digitized, then passed through an AGC stage that adaptively adjusts the signal amplitude using a square-law power estimator and accumulation-based loop to maintain dynamic range, compensating for path loss and fading. The signal is then filtered by an SRRC filter matched specifically to the transmitted pulse shape. CFO is then estimated using a known training sequence at the beginning of the packet and corrected by rotating the baseband samples with a complex exponential at frequency $-\Delta f_{\text{est}}$ using a Carrier Frequency Compensator (CFC) as discussed in Section III-B.

**Frame Detection**: The frame detection subsystem, illustrated in Fig. 4, operates immediately after coarse frequency offset correction. The frame detection and synchronization is implemented using an FIR matched filter that performs a sliding correlation between the received signal and the known Golay Complementary Sequences (GCS) embedded in the preamble. As discussed in Sec. III, the matched filter output exhibits a sharp peak at the end of the preamble due to the GCS's zero out-of-phase aperiodic autocorrelation properties. A detection flag is asserted when the peak magnitude surpasses a predefined threshold, indicating the presence of a reliable frame. Once a frame is detected, the frame synchronizer module asserts the `IS_PILOT` signal based on the location of the pilot sequence, which is inferred from the configured number of pilot repetitions per packet, $\lambda_p$.

**Channel Equalization and Baseband Demodulation**: The channel equalization subsystem, illustrated in Fig. 6, is triggered when the `IS_PILOT` flag is asserted high to estimate and compensate for channel-induced amplitude, phase, and residual frequency distortions. After channel correction, the residual phase offset can also be measured when the `IS_RESIDUAL` flag is asserted high. The corrected complex symbols are then demodulated into a byte stream, transferred to the physical layer interface via the AXI4-Stream protocol. This stream is finally forwarded to the onboard Intel NUC and passed up the protocol stack for further process.

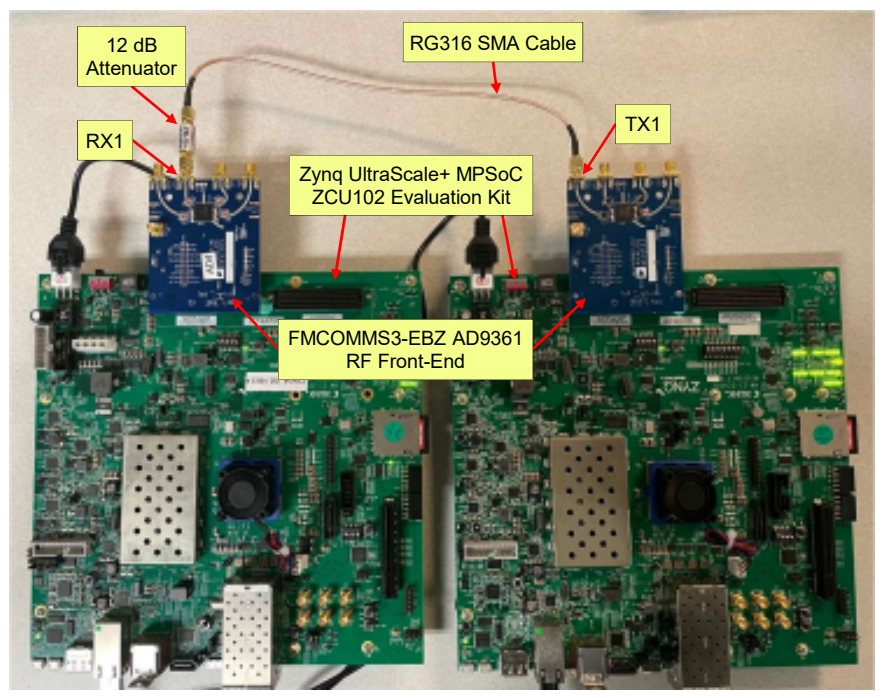

Fig. 7: Residual CFO measurement using Xilinx Zynq UltraScale+ ZCU102 MPSoC devices with wired channel.

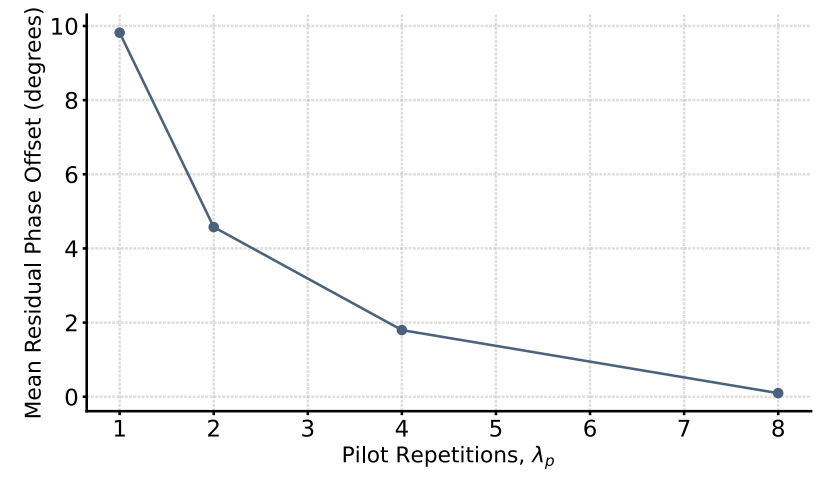

Fig. 8: Residual Phase Offset vs Pilot Repetitions.

## V. Prototyping and Experimental Evaluation

In this section, our objective is to evaluate the robustness of BenchLink through a comprehensive set of experiments in various deployment scenarios, including indoor environments, outdoor ground settings, and aerial platforms. The assessment focuses on two critical performance aspects: (i) the residual carrier frequency offset, which indicates the system's ability to maintain synchronization under challenging conditions, and (ii) the end-to-end efficiency of the link, quantified in terms of achieved goodput. Although GPS access is available on campus, it is deliberately disabled during all experiments to mimic GPS-denied environments.

### A. Indoor Emulation in Controlled Environment

Before performing outdoor OTA experiments, we first validate the effectiveness of the frequency offset compensation scheme described in Section III using an indoor testbed with emulated channels under controlled conditions. A snapshot of the setup is shown in Fig. 7. The testbed consists of two Xilinx Zynq UltraScale+ ZCU102 MPSoC platforms, each interfaced with an AD-FMCOMMS3-EBZ RF front-end transmitting and receiving radio signals at a carrier frequency of 2.485 GHz. The transmitter and receiver are connected through a SubMiniature version A (SMA) coaxial RF cable to establish a clean unidirectional communication link. The two radios operate using independent local oscillators without external synchronization, allowing us to isolate frequency drift arising from oscillator mismatch without interference from channel-induced effects. To avoid saturation at the receiver, the transmitted signal is attenuated using a 12 dB Mini-Circuits fixed RF attenuator.

TABLE I: Pilot and Data Symbols per Payload for Different Repetition Configurations

| $\lambda_p$ | Pilot Symbols | Data Symbols |
|---|---|---|
| 1 | 16 | 240 |
| 2 | 32 | 224 |
| 4 | 64 | 192 |
| 6 | 96 | 160 |
| 8 | 128 | 128 |

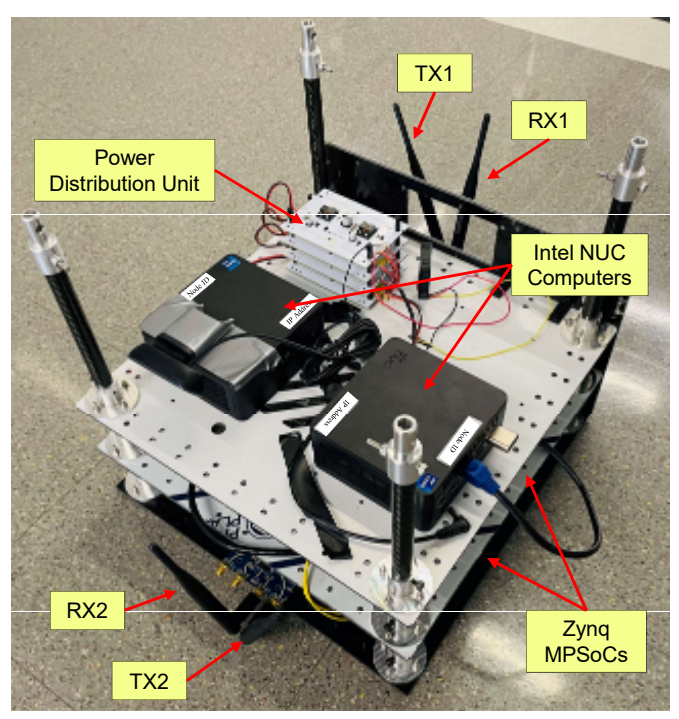

Fig. 9: Custom-designed BenchLink module with integrated Zynq UltraScale+ ZCU102 MPSoC, AD9361 RF front-end, and onboard computing units for field deployment.

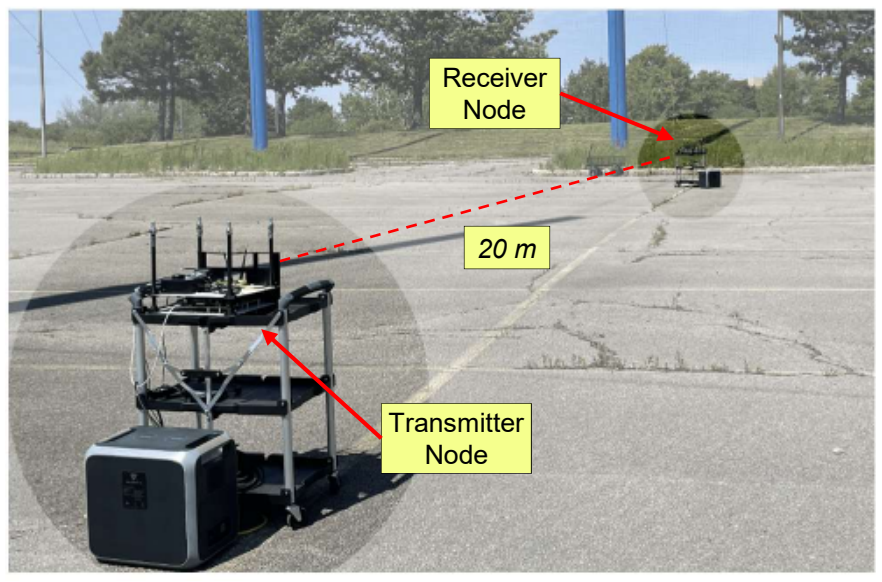

Fig. 10: Outdoor ground testbed with custom portable BenchLink modules for OTA link evaluation.

Each frame consists of a total of 256 symbols, with the distribution between the pilot and the data symbols summarized in Table I. Fig. 8 illustrates the impact of pilot repetition on the mean residual phase offset caused by clock drift. With a single pilot repetition, the offset is approximately 9.8°, which decreases to around 4.6° with two repetitions and drops below 1.8° with four repetitions. After four repetitions, the further reduction of the phase offset becomes marginal only and the offset approaches 0° at eight repetitions. These results demonstrate that, as expected, while additional pilots improve synchronization, the gain diminishes beyond a moderate level. Therefore, we consider in the following experiments a maximum of eight pilot repetitions per packet.

### B. Outdoor Ground OTA Communications

To enable field experiments with minimal deployment complexity, we developed two custom portable BenchLink modules, each integrating two radio subsystems comprising a Xil-inx Zynq UltraScale+ ZCU102 MPSoC interfaced with Analog Devices FMCOMMS3 RF front-ends programmed to transmit and receive radio signals at carrier frequency of 2.485 GHz. The MPSoC is further connected to an Intel NUC computer, as shown in Fig. 9. A snapshot of the deployed testbed is shown in Fig. 10, where each BenchLink module is mounted on a portable cart and powered by a Bluetti Solar Generator AC180 high-capacity power station. In this experiment, we evaluate five pilot repetition configurations {1, 2, 4, 6, 8} across four modulation schemes {4QAM, 8QAM, 16QAM, 64QAM}. For each combination of pilot repetition and modulation, three independent rounds of testing are conducted, each lasting six minutes, during which the webcam-captured video data is continuously transmitted. Goodput was calculated as the total number of correctly decoded packets, identified by CRC validation, multiplied by payload size (following the same configurations discussed in Section V-A) and normalized by transmission duration.

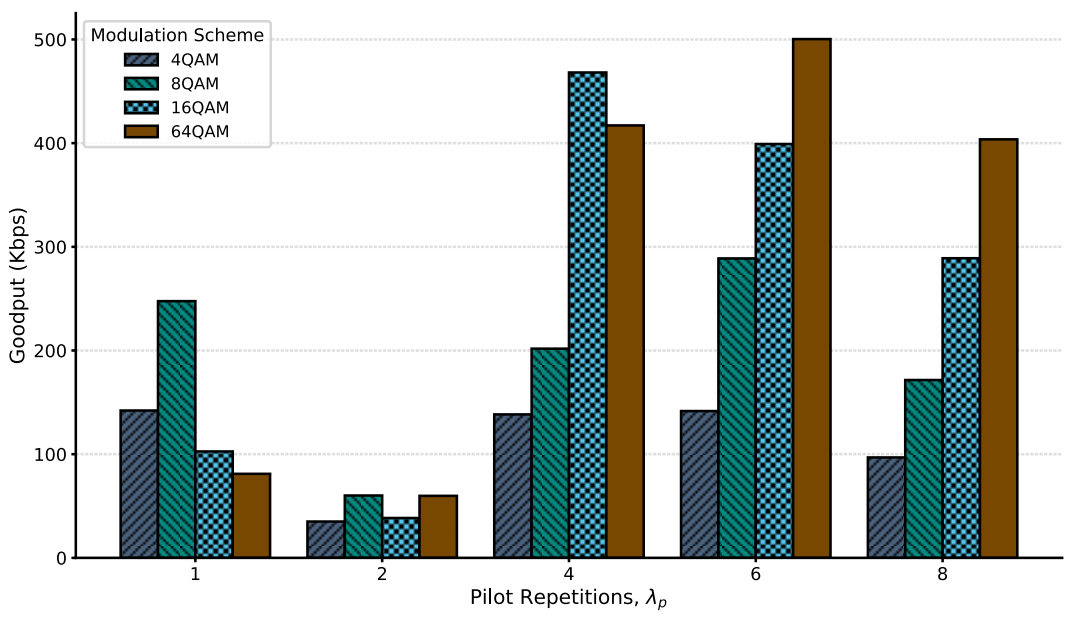

Fig. 11: Impact of pilot repetition and modulation on goodput in G2G communication link.

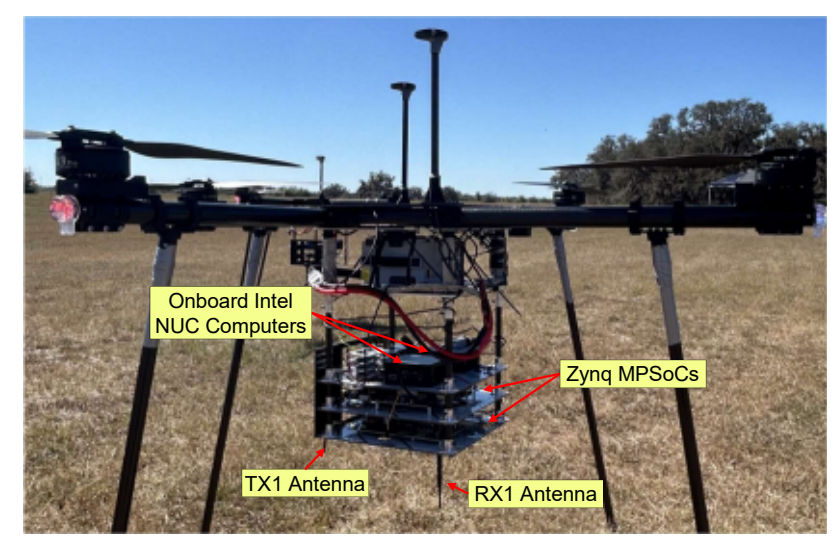

Fig. 12: Heavy-duty drone with integrated BenchLink module.

The experimental results are reported in Fig. 11. It can be seen that the goodput of the G2G link is highly sensitive to both modulation order and pilot repetition. For low-order modulation, such as 4QAM, increasing pilot repetitions has negligible impact on goodput, as the improvement in channel estimation is largely offset by pilot overhead. For 8QAM, the effect is not significant. For example, one pilot per packet achieves 247 kbps, while six repetitions raise goodput to 288 kbps, leading to an improvement of approximately 16%. The benefit of increased pilot repetitions becomes significant for higher-order modulation. For 16QAM, goodput rises from 102 kbps with one pilot to 468 kbps with four repetitions, an improvement of 356%. The goodput reaches its peak at four repetitions; after that, the pilot overhead outweighs the gain from additional channel estimates. Similarly, for 64QAM, the goodput improves from 81 kbps with one pilot to 500 kbps with six repetitions, an increase of 517%. Additional pilots beyond six again lead to reduced throughput due to overhead. From this experiment, we can learn that higher-order modulations require more frequent channel estimation and residual offset correction to maintain constellation points within their decision regions, whereas lower-order schemes are more robust and less sensitive to pilot density.

An interesting observation from this experiment is that, as shown in Fig. 11, increasing pilot repetitions from one to two consistently results in a reduction in goodput across all modulation schemes. This behavior is likely due to random channel variations combined with insufficient improvement in link quality to offset the additional pilot overhead. In the following experiments, we further investigate whether this observation is generalizable to other communication scenarios, focusing on aerial settings.

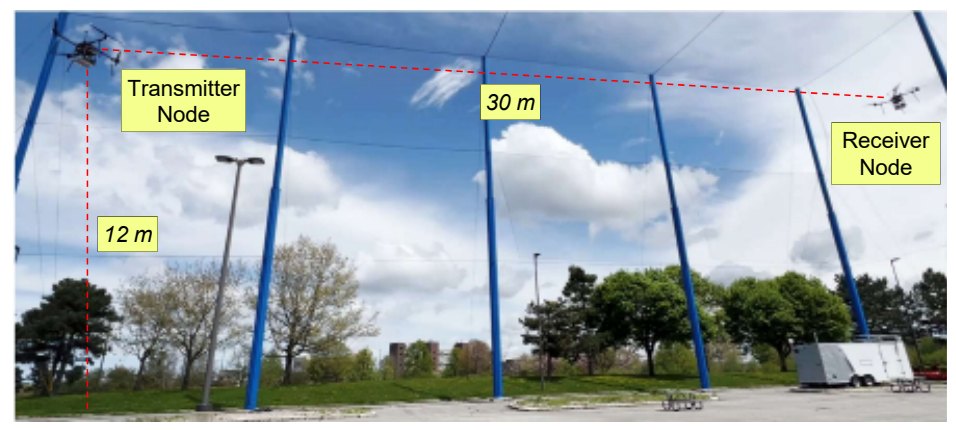

Fig. 13: OTA testing and measurement of the effects of pilot repetitions on link robustness.

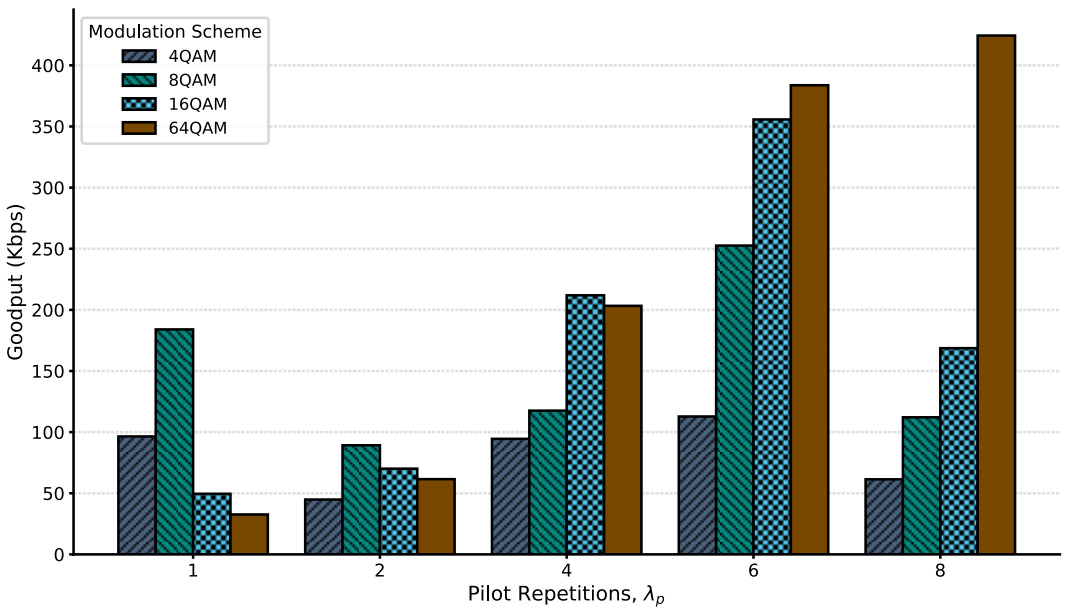

Fig. 14: Impact of pilot repetition and modulation on goodput in A2A communication links.

### C. Aerial OTA Communications

In this experiment, we further evaluate BenchLink in aerial scenarios. To this end, we custom-designed two heavy-duty drones, each capable of carrying payloads of up to 50 lbs and generating approximately 120 lbs of thrust. This capacity is sufficient to lift the BenchLink module, which weighs around 15 lbs, while providing ample thrust margin to maintain stable flight even under strong wind conditions. Fig. 12 shows a snapshot of the drone carrying the BenchLink module.

The setup of the aerial experiment is shown in Fig. 13. The flight tests were carried out at the on-campus outdoor drone testing facility, a football field-sized netted enclosure that allows UAV operations without the restrictions of FAA regulations. For safety and due to the limited flight volume inside the enclosure, the drones were operated at a fixed altitude of approximately 12 meters and maintained a stable hover throughout each experiment. The horizontal separation between the two drones was set to 30 meters. It is worth noting that operating heavy-lift drones with a takeoff weight exceeding 55 lbs requires special FAA registration and the use of a designated outdoor test site. Incorporating such extended-range testing is left for future work.

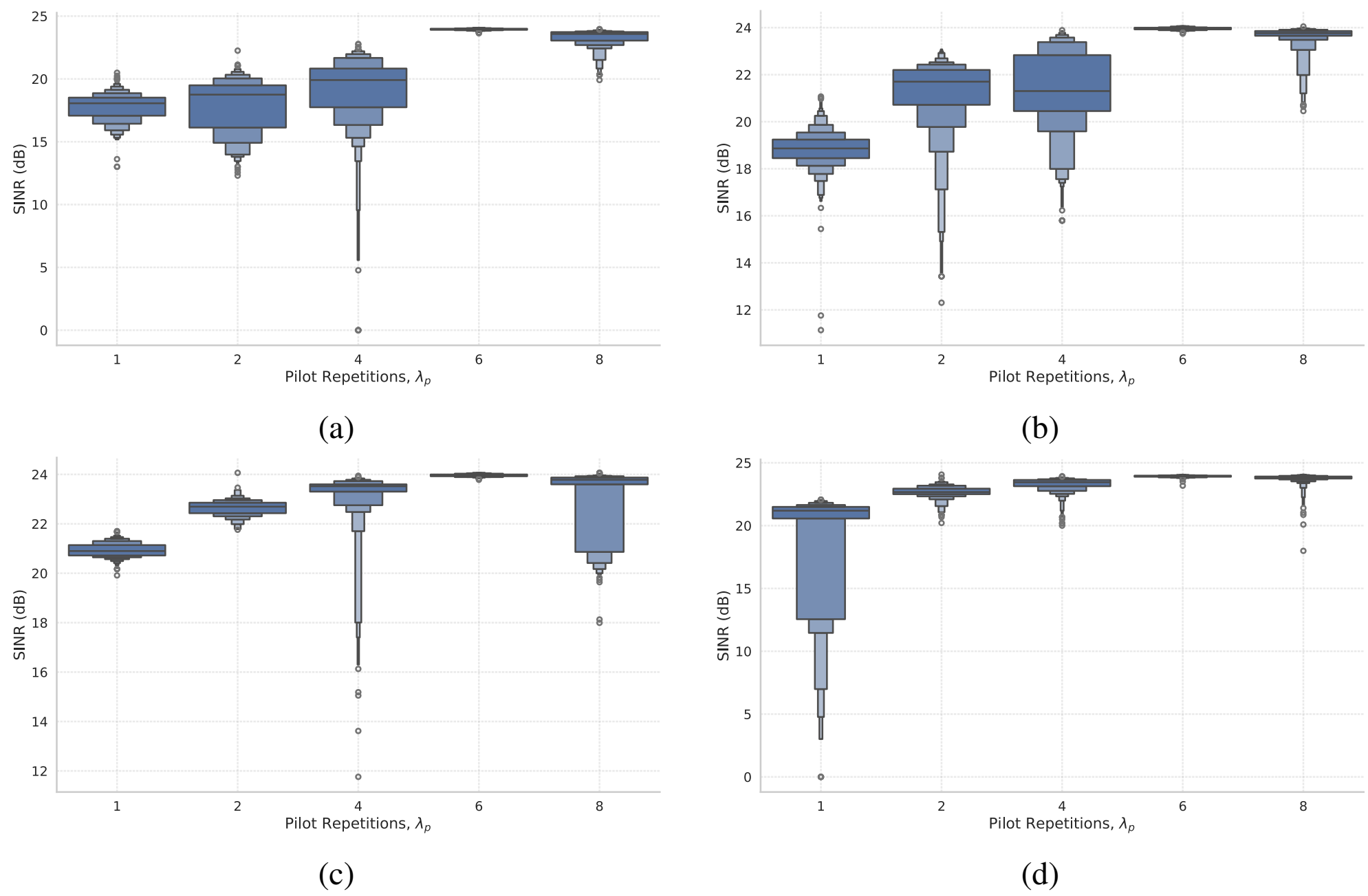

Fig. 16: Impact of $\lambda_p$ on SINR for A2A links for various modulations: (a) 4QAM; (b) 8QAM; (c) 16QAM; and (d) 64QAM.

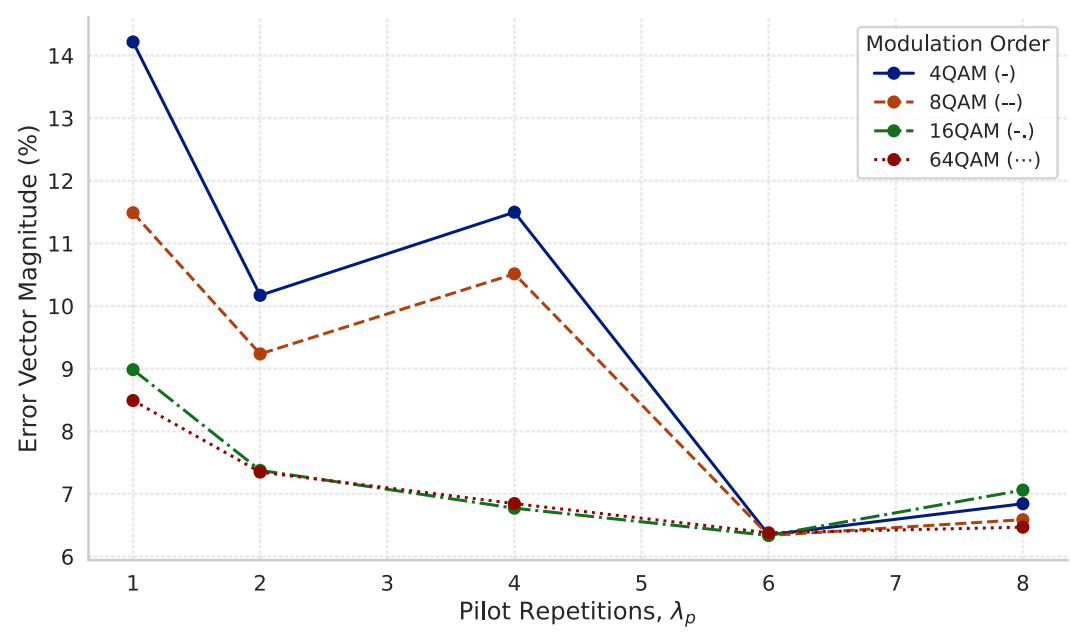

Fig. 15: EVM versus pilot repetitions for different modulation orders in A2A communication links.

Figure 14 illustrates the improvement in average goodput for A2A communication links as the number of pilot repetitions increases across different modulation schemes. For 4QAM, A2A goodput improves modestly from 96 Kbps with 1 pilot repetition to 113 Kbps with 6 repetitions, resulting in a 16% gain. For 8QAM, the goodput rises from 184 Kbps to 253 Kbps, corresponding to a 37.3% improvement. The gains become more significant with higher modulation orders: for 16QAM, goodput increases from 50 Kbps to 356 Kbps (618% improvement), while 64QAM improves from 33 Kbps to 424 Kbps, achieving a substantial 1200% gain.

**EVM and SINR Analysis.** It can be seen from Fig. 14 that a noticeable drop in goodput performance occurs at two pilot repetitions for 4QAM and 8QAM but not for 16QAM and 64QAM. To better understand this phenomenon, we further analyze the signal quality by plotting the measured Error Vector Magnitude (EVM) in Fig. 15 and SINR in Figs. 16.

EVM quantifies the deviation between the received symbols and their ideal constellation points relative to the magnitude of the ideal signal, providing a direct measure of the accuracy and distortion of the received signal [31]. For 4QAM, the EVM decreases sharply from approximately 14.2% at one pilot repetition to about 10.2% at two, then increases slightly to around 11.5% at four repetitions before dropping to 6.4% at six. A similar trend is seen in 8QAM, where the EVM drops from approximately 11.5% at one repetition to 9.2% at two, followed by a rebound to approximately 10.5% at four. In contrast, 16QAM and 64QAM exhibit smoother, monotonic decreases in EVM, starting near 9% and 8.5%, respectively, at one pilot and converging to about 6.3% at six repetitions.

The corresponding SINR results are further reported in Fig. 16. For 4QAM, shown in Fig. 16(a), average SINR increases from about 18 dB at $\lambda_p = 1$ to roughly 23 dB at $\lambda_p = 6$, with little change beyond that point. For 8QAM, shown in Fig. 16(b), the SINR rises from approximately 19 dB at $\lambda_p = 1$ to around 24 dB at $\lambda_p = 6$ before saturating. As shown in Fig. 16(c), 16QAM improves from about 21 dB at $\lambda_p = 1$ to nearly 24 dB at $\lambda_p = 6$, while the performance of 64QAM, shown in Fig. 16(d), increases from roughly 20 dB to about 24 dB over the same range. Additionally, distribution of SINR values shows that the variance significantly decreases with an increasing number of pilot repetitions.

From Figs. 15, and 16 we can see that both metrics, particularly SINR, improve smoothly and monotonically in general with the selection of $\lambda_p$ due to the line-of-sight dominant channel. However, for G2G links, all modulation schemes experience a goodput drop at $\lambda_p = 2$. The multipath-rich ground environment and small-scale fading make early-stage estimation with two pilots less effective, so the limited EVM and SINR improvement cannot compensate for the reduced payload. While A2A links can benefit more from extra pilot repetitions, a small goodput drop is still observed at two pilot repetitions for 4QAM and 8QAM. This suggests that for lower-order modulations, the modest gain in estimation accuracy between one and two pilots is insufficient to compensate for the added overhead, even in the aerial environment.

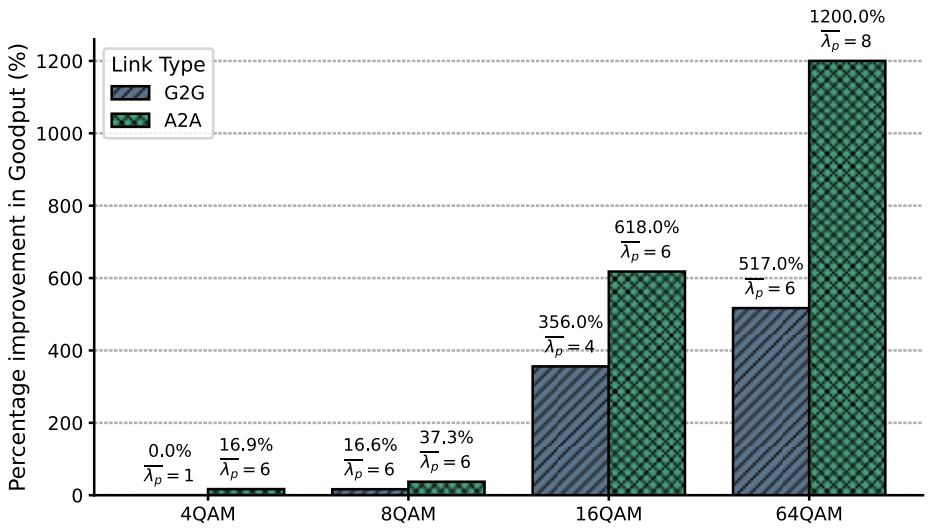

Fig. 17: Goodput improvement relative to baseline ($\lambda_p = 1$) from optimal selection of $\lambda_p$ in G2G and A2A links.

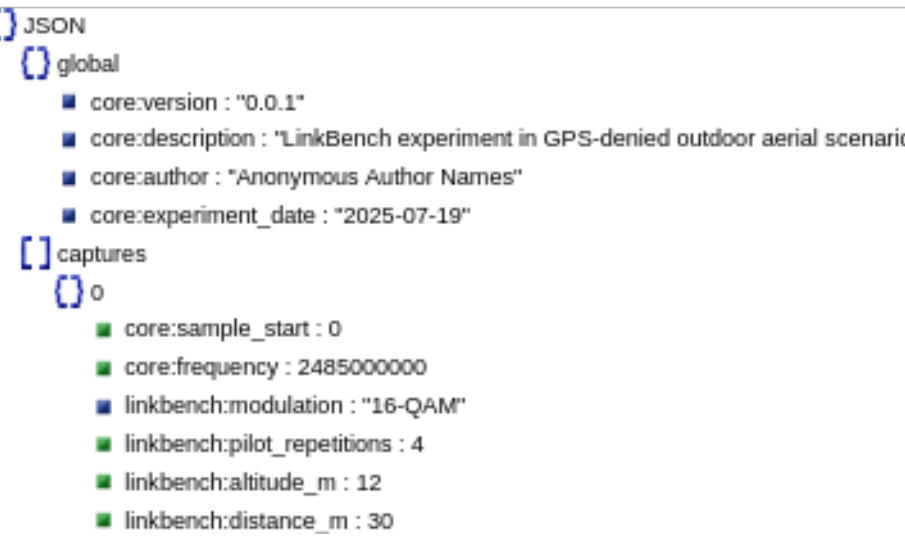

Fig. 18: Example SigMF metadata file for a BenchLink experiment.

*Summary.* As shown in Figs. 11 and 14, G2G links achieve higher absolute goodput compared to A2A UAV links. As discussed in Section III-B, this is attributed to the shorter channel coherence time inherent to UAV-enabled links. This is supported by the results shown in Fig. 17, which indicates that the relative gain from increasing $\lambda_p$ is substantially greater in A2A scenarios than in G2G scenarios. This figure reports the percentage improvement in goodput when increasing from one to $\overline{\lambda_p}$ channel corrections per packet, where $\overline{\lambda_p}$ represents the experimentally determined optimal configuration for each modulation scheme based on goodput. While G2G links show no improvement for 4QAM and only a 16.6% gain for 8QAM, A2A links achieve 16% and 37.3% respectively under the same conditions. The gap widens for higher-order modulations: 16QAM improves by 618% in A2A compared to 356% in G2G, and 64QAM achieves 1200% versus 517%.

## VI. Data Collection

We collected experimental data using the BenchLink module deployed in heterogeneous environments, including indoor, outdoor ground, and outdoor aerial scenarios. The experiments were carried out over multiple sessions to capture diverse channel conditions and operational settings. During each experiment, the BenchLink module continuously transmitted webcam-generated video payloads while logging both transmitted and received packets along with relevant metadata. The resulting dataset comprises 18 megabytes of labeled data, covering a continuous 3 hour period. Each minute contains approximately 600 data rows, amounting to a total of around 108,000 rows across the entire duration.

For each configuration, we varied key PHY layer parameters, including modulation schemes (4QAM, 8QAM, 16QAM, and 64QAM) and pilot repetitions (1, 2, 4, 6, or 8 per packet), and tested all combinations independently for two-minute intervals in burst mode. The logged data include configuration variables such as transmit and receive gains, frequency, and node altitude when applicable, as well as performance metrics such as residual frequency offset, goodput, throughput, estimated channel distribution, and EVM. All received packets were recorded with their timestamps, CRC results, and signal quality indicators.

In addition to the raw data files, we generate metadata files compliant with the Signal Metadata Format (SigMF) [32] specification. These metadata files, in JSON format, describe the parameters of each dataset, including experimental configuration, environment, link distance, and calibration details. This facilitates reproducibility and enables others to interpret and analyze the data efficiently. Fig. 18 shows a sample of the SigMF-compliant metadata accompanying each recording.

## VII. Limitations and Future Work

We believe that BenchLink provides a programmable and resilient SoC-based wireless link benchmark and can enable rigorous experimental evaluation of communication performance in GPS-denied environments. We identify several directions that will be explored in our future work.

- **Learning-based optimization:** We plan to leverage the collected dataset to train machine learning models that can dynamically determine the optimal link configuration on the fly, autonomously adapting to time-varying channel conditions.
- **Goodput enhancement:** In its current implementation, BenchLink does not employ physical-layer forward error correction (FEC) or MAC-layer automatic repeat request (ARQ), and any packet with bit errors is discarded. Future work will integrate FEC and ARQ into the protocol stack to improve goodput.
- **Large-scale testing:** Current experiments are constrained by the physical size of the netted testbed. We are establishing an outdoor testing site and pursuing collaborations with platforms such as the NSF AERPAW [33] to improve dataset diversity and channel modeling. This expansion will also allow us to evaluate network-level performance, such as scalability in multi-UAV scenarios.
- **Broader scenarios:** We aim to extend our experiments to include interference-constrained environments, such as scenarios with coexisting communications sharing the same spectrum or intentional jamming, to further validate the resilience and adaptability of the BenchLink design.

## VIII. Conclusions

In this work, we have presented BenchLink, an SoC-based benchmark for resilient wireless communication in GPS-denied environments. We designed and prototyped BenchLink on Zynq UltraScale+ MPSoCs using a hardware–software co-design architecture and evaluated BenchLink through extensive experiments in both ground and aerial scenarios to investigate the impact of pilot repetition and modulation. We also released the complete SoC-based design and dataset to the wireless community to enable research in robust UAV communications and next-generation wireless systems. In future work, we will integrate learning-based link optimization, incorporate FEC and ARQ to improve goodput, and extend testing to large-scale and interference-rich environments.

## References

[1] R. S. Cooper and A. R. Chi, "A Review of Satellite Time Transfer Technology: Accomplishments and Future Applications," *Radio Science*, vol. 14, no. 4, pp. 605 – 619, July 1979.

[2] M. Liu, R. Tu, F. Li, Q. Chen, Q. Li, J. Chen, P. Zhang, and X. Lu, "GPS + 5G Fusion for High-Precision Time Transfer," *Measurement Science and Technology*, vol. 35, no. 4, p. 045024, Jan. 2024.

[3] D. Xu, C. Liu, S. Song, and S. W. K. Ng, "Integrated Sensing and Communication in Coordinated Cellular Networks," in *Proc. of IEEE Statistical Signal Processing Workshop*, Hanoi, Vietnam, 2023.

[4] A. Mahmood, M. I. Ashraf, M. Gidlund, J. Torsner, and J. Sachs, "Time Synchronization in 5G Wireless Edge: Requirements and Solutions for Critical-MTC," *IEEE Communications Magazine*, vol. 57, no. 12, pp. 45 – 51, Dec. 2019.

[5] F. Mazzenga, R. Giuliano, and A. Vizzarri, "5G-Based Synchronous Network for Air Traffic Monitoring in Urban Air Mobility," *IEEE Access*, vol. 12, pp. 188 542 – 188 559, Dec. 2024.

[6] O. M. Picchi, F. Menzione, F. Soaulle, and J. A. D. Peral-Rosado, "Fused PNT on Future Wideband European Non-Terrestrial Network Infrastructure," in *Proc. of IEEE/ION Position, Location and Navigation Symposium (PLANS)*, Salt Lake City, UT, USA, Apr. 2025.

[7] F. Alrefaei, A. Alzahrani, H. Song, and S. Alrefaei, "A Survey on the Jamming and Spoofing Attacks on the Unmanned Aerial Vehicle Networks," in *Proc. of IEEE International IOT, Electronics and Mechatronics Conference (IEMTRONICS)*, Toronto, ON, Canada, June 2022.

[8] I. Srisomboon and S. Lee, "Positioning and Navigation Approaches Using Packet Loss-Based Multilateration for UAVs in GPS-Denied Environments," *IEEE Access*, vol. 12, pp. 13 355 – 13 369, Jan. 2024.

[9] L. Zhang, H. Zhao, J. Chen, L. Li, and X. Liu, "Vehicular Positioning Based on GPS/IMU Data Fusion Aided by V2X Networks," *IEEE Sensors Journal*, vol. 24, no. 6, pp. 9032 – 9043, Mar. 2024.

[10] S. Li, R. Tao, M. Zhang, Y. Chen, X. Xu, and X. Chu, "Positioning and Synchronization of UAVs Swarm for Distributed Coherent Signal Reception," in *Proc. of IEEE International Conference on Unmanned Systems (ICUS)*, Hefei, China, Oct. 2023.

[11] X. Jin, J. An, C. Du, G. Pan, S. Wang, and D. Niyato, "Frequency-Offset Information Aided Self Time Synchronization Scheme for High-Dynamic Multi-UAV Networks," *IEEE Transactions on Wireless Communications*, vol. 23, no. 1, pp. 607 – 620, Jan. 2024.

[12] K. F. Hasan, C. Wang, Y. Fend, and Y. Tian, "Time Synchronization in Vehicular Ad-Hoc Networks: A Survey on Theory and Practice," *Vehicular Communications*, vol. 14, pp. 39 – 51, Oct. 2018.

[13] A. Tewari, S. S. Jha, A. Sneh, S. J. Darak, and S. S. Ram, "Reconfigurable Radar Signal Processing Accelerator for Integrated Sensing and Communication System," *IEEE Transactions on Aerospace and Electronic Systems*, vol. 61, no. 1, pp. 162 – 181, Feb. 2024.

[14] D. Brunner, L. G. de Oliveira, C. Muth, S. Mandelli, M. Henninger, and A. Diewald, "Bistatic OFDM-Based ISAC With Over-the-Air Synchronization: System Concept and Performance Analysis," *IEEE Transactions on Microwave Theory and Techniques*, vol. 73, no. 5, pp. 3016–3029, May 2025.

[15] V. Savaux, C. Delacourt, and P. Savelli, "On Time-Frequency Synchronization in LoRa System: From Analysis to Near-Optimal Algorithm," *IEEE Internet of Things Journal*, vol. 9, no. 12, pp. 10 200 – 10 211, June 2022.

[16] T. Liu, Y. Hu, Y. Hua, and H. Jiang, "Study on Autonomous and Distributed Time Synchronization Method for Formation UAVs," in *Proc. of Joint Conference of the IEEE International Frequency Control Symposium & the European Frequency and Time Forum*, Denver, CO, USA, Apr. 2015.

[17] N. Mastronarde, D. Russell, Z. Guan, G. Sklivanitis, E. S. Bentley, and M. Medley, "RF-SITL: A Software-in-the-loop Channel Emulator for UAV Swarm Networks," in *Proc. of IEEE WoWMoM Workshop on Wireless Networking, Planning, and Computing for UAV Swarms (SwarmNet)*, Belfast, United Kingdom, June 2022.

[18] A. Dey, M. McManus, J. Z. Zhang, G. Sun, N. Mastronarde, E. S. Bentley, and Z. Guan, "AirTwinX: A High-Fidelity Digital Twin for Advanced Air Mobility with Ray Tracing," in *Proc. of IEEE CCNC*, Las Vegas, NV, January 2025.

[19] S. Santhinivas, P. S. P. V. Kumar, M. McManus, H. Nouri, G. Sklivanitis, D. Pados, E. S. Bentley, N. Mastronarde, and Z. Guan, "Resilient Communications with Lightweight Signature Synchronization on MPSoC Radios," in *Proc. of IEEE CCNC*, Las Vegas, NV, January 2025.

[20] M. McManus, Y. Cui, Z. J. Zhang, E. S. Bentley, M. Medley, N. Mastronarde, and Z. Guan, "On the Effects of Modeling on the Sim-to-Real Transfer Gap in Twinning the POWDER Platform," in *Proc. of IEEE GLOBECOM Workshop on Testbeds as a Service for Future Networks: Challenges & State of the Art*, Cape Town, South Africa, Dec. 2024.

[21] M. McManus, Y. Cui, J. Zhang, J. Hu, S. K. Moorthy, N. Mastronarde, E. S. Bentley, M. Medley, and Z. Guan, "Digital Twin-Enabled Domain Adaptation for Zero-Touch UAV Networks: Survey and Challenges," *Elsevier Journal of Computer Networks*, November 2023.

[22] M. McManus, Z. Guan, N. Mastronarde, and S. Zou, "On the Source-to-Target Gap of Robust Double Deep Q-Learning in Digital Twin-Enabled Wireless Networks," in *Proc. of SPIE Conference Big Data IV: Learning, Analytics, and Applications*, Orlando, Florida, April 2022, pp. 1–12.

[23] H. den Boer, R. W. D. Muller, S. Wong, and V. Voogt, "FPGA-based Deep Learning Accelerator for RF Applications," in *Proc. of IEEE MILCOM*, San Diego, CA, USA, Dec. 2021.

[24] M. Pini, F. Corraro, G. Marucco, R. Senatore, M. Perlmutter, and A. Pizzarulli, "Tightly Coupled Integration as a Strategy for Robust Navigation: Civitanavi Systems' Approach," in *Proc. of DGON Inertial Sensors and Applications (ISA)*, Braunschweig, Germany, Oct. 2024.

[25] Analog Devices, "libiio." [Online]. Available: https://github.com/analogdevicesinc/libiio

[26] Ettus Research, "Knowledge base," 2025. [Online]. Available: https://kb.ettus.com/Knowledge_Base

[27] A. Goldsmith, *Wireless Communications*. Cambridge, UK: Cambridge University Press, 2005.

[28] M. Banagar and H. Dhillon, "Wobbling and Impairments-Aware Channel Model and its Implications on High-Frequency UAV Links," in *Proc. of IEEE GLOBECOM*, Rio de Janeiro, Brazil, Dec. 2022.

[29] Y. Dou, Z. Lian, Y. Wang, B. Zhang, H. Luo, and C. Zu, "Channel Modeling and Performance Analysis of UAV-Enabled Communications With UAV Wobble," *IEEE Communications Letters*, vol. 28, no. 12, pp. 2749–2753, Oct. 2024.

[30] S. K. Moorthy and Z. Guan, "Beam Learning in MmWave/THz-band Drone Networks Under In-Flight Mobility Uncertainties," *IEEE Transactions on Mobile Computing*, vol. 21, no. 6, pp. 1945–1957, June 2022.

[31] H. Cho, J. Lee, K. Shin, and J. Yu, "EVM Analysis of Chip-Integrated Phased Array Antenna Using USRP for 5G Communication," in *Proc. of European Microwave Conference (EuMC)*, Paris, France, Sept. 2024.

[32] B. Hilburn, N. West, T. O'Shea, and T. Roy, "SigMF: The Signal Metadata Format," in *Proc. of the 8th Annual GNU Radio Conference*, Henderson, NV, USA, Sept. 2022.

[33] V. Marojevic, I. Guvenc, R. Dutta, M. L. Sichitiu, and B. Floyd, "Advanced Wireless for Unmanned Aerial Systems: 5G Standardization, Research Challenges, and AERPAW Architecture," *IEEE Vehicular Technology Magazine*, vol. 15, no. 2, pp. 22–30, June 2020.